\documentclass[twocolumn,showpacs,preprintnumbers,amsmath,amssymb,aps,prl,superscriptaddress]{revtex4}
\usepackage{color}
\usepackage{graphicx}% Include figure files
\usepackage{dcolumn}% Align table columns on decimal point
\usepackage{bm}% bold math

\def\Journal#1#2#3#4{{#1}\, {\bf #2}, #3 (#4)}
  
\def\NPA{Nucl.\ Phys.\ A}
\def\NPB{Nucl.\,Phys.\,B}
\def\PLB{Phys.\ Lett.\ B}
\def\PRL{Phys.\ Rev.\ Lett.}
\def\PRD{Phys.\,Rev.\,D}
\def\PRC{Phys.\,Rev.\,C}

\def\xbj{x}

\begin{document}

\preprint{}

\title{Forward Neutral Pion Production in p+p and d+Au Collisions 
at $\sqrt{s_{NN}}=200\,$GeV}

\affiliation{Argonne National Laboratory, Argonne, Illinois 60439}
\affiliation{University of Birmingham, Birmingham, United Kingdom}
\affiliation{Brookhaven National Laboratory, Upton, New York 11973}
\affiliation{California Institute of Technology, Pasadena, California 91125}
\affiliation{University of California, Berkeley, California 94720}
\affiliation{University of California, Davis, California 95616}
\affiliation{University of California, Los Angeles, California 90095}
\affiliation{Carnegie Mellon University, Pittsburgh, Pennsylvania 15213}
\affiliation{Creighton University, Omaha, Nebraska 68178}
\affiliation{Nuclear Physics Institute AS CR, 250 68 \v{R}e\v{z}/Prague, Czech Republic}
\affiliation{Laboratory for High Energy (JINR), Dubna, Russia}
\affiliation{Particle Physics Laboratory (JINR), Dubna, Russia}
\affiliation{University of Frankfurt, Frankfurt, Germany}
\affiliation{Institute of Physics, Bhubaneswar 751005, India}
\affiliation{Indian Institute of Technology, Mumbai, India}
\affiliation{Indiana University, Bloomington, Indiana 47408}
\affiliation{Institut de Recherches Subatomiques, Strasbourg, France}
\affiliation{University of Jammu, Jammu 180001, India}
\affiliation{Kent State University, Kent, Ohio 44242}
\affiliation{Lawrence Berkeley National Laboratory, Berkeley, California 94720}
\affiliation{Massachusetts Institute of Technology, Cambridge, MA 02139-4307}
\affiliation{Max-Planck-Institut f\"ur Physik, Munich, Germany}
\affiliation{Michigan State University, East Lansing, Michigan 48824}
\affiliation{Moscow Engineering Physics Institute, Moscow Russia}
\affiliation{City College of New York, New York City, New York 10031}
\affiliation{NIKHEF and Utrecht University, Amsterdam, The Netherlands}
\affiliation{Ohio State University, Columbus, Ohio 43210}
\affiliation{Panjab University, Chandigarh 160014, India}
\affiliation{Pennsylvania State University, University Park, Pennsylvania 16802}
\affiliation{Institute of High Energy Physics, Protvino, Russia}
\affiliation{Purdue University, West Lafayette, Indiana 47907}
\affiliation{Pusan National University, Pusan, Republic of Korea}
\affiliation{University of Rajasthan, Jaipur 302004, India}
\affiliation{Rice University, Houston, Texas 77251}
\affiliation{Universidade de Sao Paulo, Sao Paulo, Brazil}
\affiliation{University of Science \& Technology of China, Hefei 230026, China}
\affiliation{Shanghai Institute of Applied Physics, Shanghai 201800, China}
\affiliation{SUBATECH, Nantes, France}
\affiliation{Texas A\&M University, College Station, Texas 77843}
\affiliation{University of Texas, Austin, Texas 78712}
\affiliation{Tsinghua University, Beijing 100084, China}
\affiliation{Valparaiso University, Valparaiso, Indiana 46383}
\affiliation{Variable Energy Cyclotron Centre, Kolkata 700064, India}
\affiliation{Warsaw University of Technology, Warsaw, Poland}
\affiliation{University of Washington, Seattle, Washington 98195}
\affiliation{Wayne State University, Detroit, Michigan 48201}
\affiliation{Institute of Particle Physics, CCNU (HZNU), Wuhan 430079, China}
\affiliation{Yale University, New Haven, Connecticut 06520}
\affiliation{University of Zagreb, Zagreb, HR-10002, Croatia}

\author{J.~Adams}\affiliation{University of Birmingham, Birmingham, United Kingdom}
\author{M.M.~Aggarwal}\affiliation{Panjab University, Chandigarh 160014, India}
\author{Z.~Ahammed}\affiliation{Variable Energy Cyclotron Centre, Kolkata 700064, India}
\author{J.~Amonett}\affiliation{Kent State University, Kent, Ohio 44242}
\author{B.D.~Anderson}\affiliation{Kent State University, Kent, Ohio 44242}
\author{D.~Arkhipkin}\affiliation{Particle Physics Laboratory (JINR), Dubna, Russia}
\author{G.S.~Averichev}\affiliation{Laboratory for High Energy (JINR), Dubna, Russia}
\author{S.K.~Badyal}\affiliation{University of Jammu, Jammu 180001, India}
\author{Y.~Bai}\affiliation{NIKHEF and Utrecht University, Amsterdam, The Netherlands}
\author{J.~Balewski}\affiliation{Indiana University, Bloomington, Indiana 47408}
% ...

\collaboration{STAR Collaboration}\homepage{www.star.bnl.gov}\noaffiliation

\date{\today}

\begin{abstract}
Measurements of the production of forward $\pi^0$ mesons from 
p+p and d+Au collisions at $\sqrt{s_{NN}}=200\,$GeV are 
reported.
The p+p yield generally agrees with next-to-leading order 
perturbative QCD calculations.
The d+Au yield per binary collision
is suppressed as $\eta$ increases, decreasing to $\sim 30\%$ of the 
p+p yield at $\langle\eta\rangle=4.00$, well below
shadowing expectations.
Exploratory measurements of azimuthal correlations of the 
forward $\pi^0$ with charged hadrons at $\eta\approx 0$ show a 
recoil peak in p+p that is suppressed in d+Au at low 
pion energy.
These observations are qualitatively consistent with a 
saturation picture of the low-$\xbj$ gluon structure of heavy nuclei.
\end{abstract}

\maketitle

Little is known about the gluon structure of heavy nuclei 
\cite{hirai}.
For protons, the gluon parton distribution function (g-PDF) is 
constrained at small $\xbj$ (fraction of nucleon momentum)
primarily by scaling violations observed in 
deep-inelastic lepton scattering (DIS) at the HERA collider 
\cite{hera}.
The proton DIS data are accurately described by evolution equations of 
Quantum Chromodynamics (QCD) that allow the determination of the 
g-PDF \cite{evolution}.
As $\xbj$ decreases, the g-PDF is found to increase from gluon 
splitting as the partons evolve.
At a sufficiently small value of $\xbj$, yet to be determined by 
experiment, the splitting is expected to 
become balanced by recombination as the gluons overlap, resulting 
in gluon saturation \cite{saturation}.
At a given $\xbj$, 
the density of gluons per unit transverse area is expected to be 
larger in nuclei than in nucleons, thus, nuclei provide a natural
environment in which to search for gluon saturation.
Fixed target nuclear DIS experiments are restricted in the 
kinematics available; they have determined the nuclear g-PDF 
only for $\xbj\stackrel{>}{_\sim}0.02$ \cite{hirai}.

Using factorization in a perturbative QCD (pQCD) framework, 
PDFs and fragmentation functions (FFs) measured in 
electromagnetic reactions are used to calculate hadronic 
processes.
In p+p collisions, Next-to-Leading Order (NLO) pQCD calculations  
quantitatively describe inclusive $\pi^0$ production over a broad 
range of pseudorapidity ($\eta=-\ln[\tan(\theta/2)]$)
at center-of-mass energy $\sqrt{s}=200\,$GeV
\cite{STARpi0,PHENIXpi0}, but not at lower $\sqrt{s}$ 
\cite{soffer}.
In pQCD, hadroproduction at large $\eta$ from p+p collisions 
at $\sqrt{s}=200\,$GeV probes gluons in one proton using the 
valence quarks of the other, covering a broad 
distribution of gluon $\xbj$ peaked around 0.02 \cite{gsv}.
Analogously, hadroproduction in the d-beam
(forward) direction of d+Au collisions 
is sensitive to the gluon structure of the Au nucleus.
Quantifying if saturation occurs at RHIC energies is 
important because the matter created in heavy-ion 
collisions comes predominantly from the collisions of low-$\xbj$ 
gluons \cite{gyulassymclerran}.
Recently, the yield of forward negatively charged hadrons
($h^-$) in d+Au collisions was found to be suppressed 
relative to p+p \cite{BRAHMS}.
The suppression is especially significant since isospin effects 
should reduce $h^-$ production in p+p collisions, but 
not in d+Au \cite{gsv}.
 
Many models try to describe forward hadroproduction from 
heavy nuclei.
In the Color Glass Condensate (CGC) formulation, the 
low-$\xbj$ gluon density is saturated, resulting in dense color fields
that scatter the partons from the deuteron beam \cite{cgc}.
The average gluon-$\xbj$ decreases rapidly with
increasing $\eta$ to $\approx 10^{-4}$ for pions produced at $\eta=4$ 
\cite{cgc-cross-section}.
Another approach scatters quarks coherently from multiple
nucleons, leading to an effective shift in gluon-$\xbj$
\cite{coherent}.
Shadowing models modify the nuclear g-PDF in a standard 
factorization framework \cite{gsv,shadowing}.
Other models include 
limiting fragmentation \cite{limitingfrag},
parton recombination \cite{recombination}, and factorization breaking 
\cite{factorization}.

Additional insight into the particle production mechanism can be 
gained by analyzing the azimuthal correlations ($\Delta\phi$) of 
the forward $\pi^0$ with coincident hadrons.
Assuming collinear elastic parton ($2\rightarrow 2$) scattering, 
a back-to-back peak at $\Delta\phi=\pi$ is expected, 
with the rapidity of the recoil particle correlated with 
$\xbj$ of the struck gluon.
In a saturation picture, the quark undergoes multiple 
interactions through the dense gluon field, resulting in 
multiple recoil partons instead of a single one
\cite{coherent,monojet}, thereby modifying the $\Delta\phi$ 
distribution and possibly leading to the appearance of 
monojets \cite{kharzeevmonojet}.
  
\begin{figure}
\includegraphics[height=2.7in]{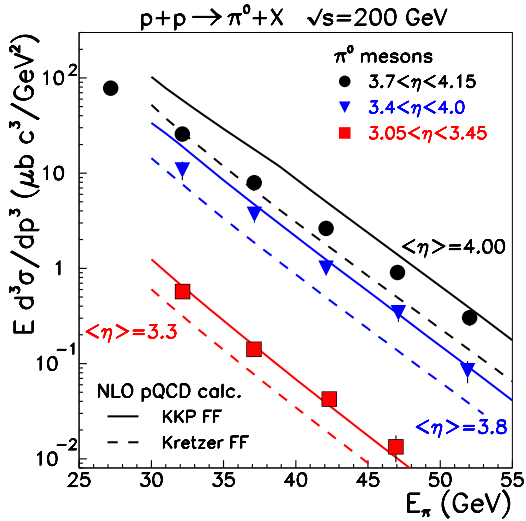}
\caption{
Inclusive $\pi^0$ cross section for p+p collisions
versus the leading $\pi^0$ energy ($E_\pi$) averaged over 5 GeV bins 
at fixed 
pseudorapidity ($\eta$).
The error bars combine 
statistical and point-to-point systematic errors.
The curves are NLO pQCD calculations using two sets of
fragmentation functions (FF).
\label{fig:inclusive}}
\end{figure}

\begin{figure}
\includegraphics[height=2.7in]{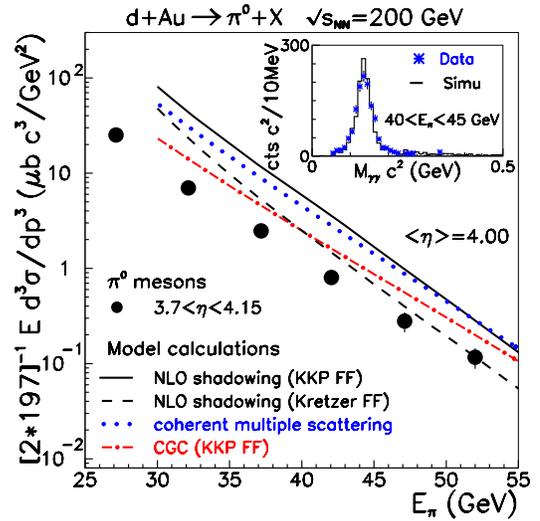}
\caption{
Inclusive $\pi^0$ cross section per binary 
collision
for d+Au collisions, as in Fig.~\ref{fig:inclusive}.
The curves are calculations described in the text.
(Inset) Diphoton invariant mass 
spectrum for data 
(stars), normalized to simulation (histogram).  
\label{fig:dau}}
\end{figure}

In this Letter, we present the yields of high energy 
$\pi^0$ mesons ($25<E_\pi<55\,$GeV) at forward rapidities 
($3.0\stackrel{<}{_\sim}\eta\stackrel{<}{_\sim}4.2$) from p+p 
(Fig.~\ref{fig:inclusive}) and 
d+Au (Fig.~\ref{fig:dau}) collisions at $\sqrt{s_{NN}}=200\,$GeV.
The data are compared with models and with
$h^-$ data at smaller $\eta$.
The $\Delta\phi$ distributions of the forward $\pi^0$ 
with midrapidity $h^\pm$ are presented.
                                                                        
Data were collected by the STAR experiment (Solenoid Tracker at
RHIC) at the Brookhaven National Laboratory Relativistic Heavy Ion
Collider (RHIC). 
At midrapidity, a time projection chamber is used to 
detect charged particles, while a forward $\pi^0$ detector (FPD) 
is used at forward rapidities.
In 2002, p+p collisions were studied with a prototype FPD 
(PFPD) \cite{STARpi0}.
In 2003, p+p collisions were studied with the complete FPD
and exploratory measurements were made for d+Au collisions.

The luminosity was determined using the rate of coincidences 
on either side of the interaction region between beam-beam 
counters (BBC) for p+p collisions \cite{STARpi0} and 
zero-degree calorimeters for d+Au collisions \cite{STARdau}.
For p+p, the transverse size of the colliding beams and the
number of colliding ions were measured, giving a coincidence
cross section of 
$26.1\pm 0.2\,({\rm stat})\pm 1.8\,({\rm sys})\,$mb
\cite{drees}.
For d+Au, the coincidence cross section was measured to
be $(19.2\pm 1.3)\%$ of the hadronic cross section,
$\sigma^{dAu}_{\rm hadr}$ \cite{STARdau}.
The integrated luminosity for these data was 
$\approx 350\,$nb$^{-1}$ ($200\,\mu{\rm b}^{-1}$) for p+p (d+Au) 
collisions.

Events required more energy in the calorimeter
than from a 15-GeV electron. 
A BBC coincidence reduces non-collision background 
but requires an $E_\pi$-independent $10\%$ correction to the 
yields \cite{STARpi0} to account for its efficiency.
The energy is calibrated to $\approx 1\%$ from the centroid of 
the $\pi^0$ peak in the diphoton invariant mass, 
$M_{\gamma\gamma}$ \cite{moriond05}.
Monte Carlo simulations with physics backgrounds 
and the full detector response describe p+p and d+Au data 
for many variables, e.\,g., $M_{\gamma\gamma}$ in 
Fig.~\ref{fig:dau} (inset).
Jet background is reduced in the FPD by requiring 
two reconstructed photons ($N_\gamma =2$), selecting 
$78\%\ (53\%)$ of events with $E_\pi>25\,$ GeV and 
$N_\gamma\ge 2$ in p+p (d+Au) data. 
The $\pi^0$ detection efficiency is determined in a matrix 
of $E_\pi$ and $\eta$ from background-corrected simulations.
For d+Au it is dominated by the FPD geometrical acceptance
and is within 8-19\% of the efficiency in p+p.

Inclusive $\pi^0$ cross sections for p+p 
collisions at $\sqrt{s}=200\,$GeV are seen in 
Fig.~\ref{fig:inclusive} at 
$\langle\eta\rangle =3.3$, 3.8 \cite{STARpi0}, and 4.00.
Data are in 5 GeV bins, plotted at the 
average $E_\pi$.
Data at $\langle\eta\rangle = 3.3$ and 3.8 were taken with 
the PFPD, where the systematic error increases with $E_\pi$ from 
$10-26\%$, dominated by the correction for the jet accompanying 
the $\pi^0$ \cite{STARpi0}.
Data at $\langle\eta\rangle=4.00$ were taken with the FPD, where 
the systematic error is $8-16\%$, dominated by the energy 
calibration \cite{moriond05}.
The normalization error is 17\% for both p+p and d+Au, 
dominated by the absolute $\eta$ uncertainty \cite{moriond05}.
The curves are NLO pQCD calculations 
\cite{aversa} 
using CTEQ6M PDFs \cite{cteq} and equal 
renormalization and factorization scales of 
$p_T = E_\pi/\cosh\eta$.
Scale dependence is comparable at
$\eta\approx 4$ and $\eta\approx 0$.
Theoretical systematic errors, attributed to 
scale dependence at $\eta\approx 0$ \cite{PHENIXpi0}, may 
require further study at large $\eta$.
The solid and dashed curves use Kniehl-Kramer-P\"{o}tter (KKP) 
\cite{kkp} and 
Kretzer \cite{kretzerff} FFs, respectively,
which differ primarily in the gluon-to-pion FF. 
Differences between FFs may occur at 
$p_T\stackrel{<}{_\sim}2\,$GeV/c, where 
the dominant contribution to $\pi^0$ production becomes 
gg scattering \cite{kretzer}.
At $\langle\eta\rangle=3.3$ and 3.8, the data are consistent 
with KKP.
At $\langle\eta\rangle=4.00$, the data drop below 
KKP and approach Kretzer as $p_T$ decreases, 
similar to the trend seen at $\eta\approx 0$ \cite{PHENIXpi0}.

The study of effects from possible gluon saturation in a 
nucleus begins with the inclusive $\pi^0$ cross 
section for d+Au collisions (Fig.~\ref{fig:dau}). 
No explicit constraint is placed on the centrality of the 
collisions analyzed.
The systematic error is $10-22\%$, dominated by the background 
correction.
The solid (dashed) curve is a NLO pQCD 
calculation using Au PDFs with shadowing \cite{gsv} and
KKP (Kretzer) FFs.
The dotted curve is a LO calculation of multiple parton 
scattering \cite{coherent}, normalized to $\pi^0$ data at 
$\eta\approx 0$ \cite{PHENIXpi0}.
The dot-dash curve is a LO calculation convoluting CTEQ5 PDFs 
and KKP FFs, replacing the hard partonic scattering with a 
dipole-nucleus cross section to model parton scattering from a 
CGC in the nucleus \cite{cgc-cross-section}, normalized to 
$d+Au\rightarrow h^- +X$ data at $\eta=3.2$ \cite{BRAHMS}.  
The CGC calculation overpredicts the $\pi^0$ data here by a 
factor of 2, a factor that could approach unity with use of 
the Kretzer FF.
The $p_T$ dependence of the yield is consistent with the CGC 
calculation.
                                                                               
\begin{figure}
\includegraphics[height=2.65in]{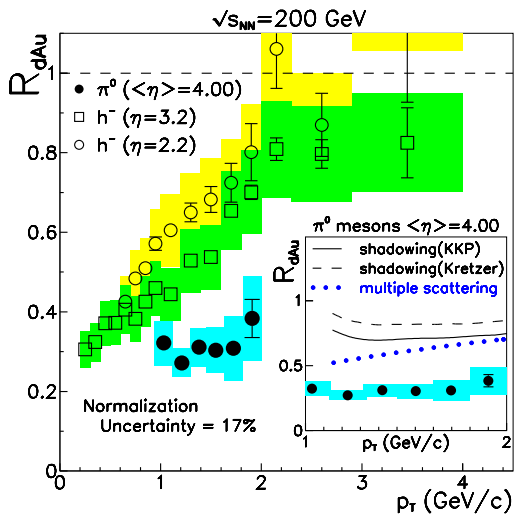}
\caption{
Nuclear modification factor ($R_{\rm dAu}$) for minimum-bias 
d+Au collisions versus transverse momentum ($p_T$).
The solid circles are for $\pi^0$ mesons.
The open circles and boxes are for negative hadrons 
\cite{BRAHMS}.
The error bars are statistical, while the shaded boxes are 
point-to-point systematic errors.
(Inset) $R_{\rm dAu}$ for $\pi^0$ mesons 
with the ratio of 
curves in Figs.~\ref{fig:dau} and
\ref{fig:inclusive}.
\label{fig:rda}}
\end{figure}

The nuclear modification factor is defined as:
\begin{equation}
R^Y_{\rm dAu} = \frac{ \sigma_{\rm inel}^{pp} }
                 { \langle N_{\rm bin} \rangle
                   \sigma_{\rm hadr}^{dAu}}
            \frac{ E\,d^3\sigma/dp^3 (d+Au\rightarrow Y+X)}
                 { E\,d^3\sigma/dp^3 (p+p\rightarrow Y+X)}.
\end{equation}
The inelastic p+p cross section is 
$\sigma_{\rm inel}^{pp}=42\,$mb, while 
$\sigma_{\rm hadr}^{dAu}=(2.21\pm0.09)\,$b and the mean
number of binary collisions,
$\langle N_{\rm bin}\rangle=7.5\pm 0.4$, are from a
Glauber model calculation \cite{STARdau}.
The prefactor in $R^Y_{\rm dAu}$ is equal to the ratio of 
binary collisions in p+p and d+Au, $1/(2\times 197)$.
Fig.~\ref{fig:rda} shows $R^{\pi^0}_{\rm dAu}$
versus $p_T$ at $\langle\eta\rangle=4.00$ with $h^-$ 
data at smaller $\eta$ \cite{BRAHMS}.
Systematic errors from p+p and d+Au are added in 
quadrature.
The normalization error includes the $\langle N_{\rm bin}\rangle$
error but not the absolute $\eta$ error, 
since the FPD position was the same for d+Au and p+p data.

In the absence of nuclear effects, hard processes 
scale with the number of binary collisions and $R^Y_{\rm dAu}=1$.
At midrapidity, $R^{\,h^\pm}_{\rm dAu}\stackrel{>}{_\sim}1$,
with a Cronin enhancement for 
$p_T\stackrel{>}{_\sim}2\,$GeV/c \cite{BRAHMS,STARdau}.
As $\eta$ increases, $R^Y_{\rm dAu}$ becomes much less than 1.
This decrease with $\eta$ is qualitatively
consistent with models that suppress the nuclear gluon
density
\cite{cgc,coherent,shadowing,recombination}.
Scaling $R^{\,h^-}_{\rm dAu}$ 
by 2/3 to account for isospin effects on 
$p+p\rightarrow h^-+X$ 
\cite{gsv}, $R^{\,\pi^0}_{\rm dAu}$ is consistent with a 
linear extrapolation of the scaled $R^{\,h^-}_{\rm dAu}$ to 
$\eta =4$.
The curves in the inset are ratios of the 
calculations in Figs.~\ref{fig:dau} and 
\ref{fig:inclusive}.
The data lie below all the predictions.

\begin{figure}
\includegraphics[height=2.65in]{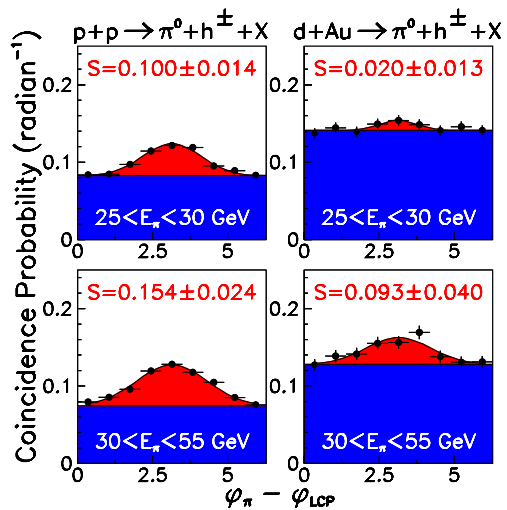}
\caption{
Coincidence probability versus azimuthal angle difference between 
the forward $\pi^0$ and a leading charged particle at 
midrapidity with $p_T>0.5\,$GeV/c.
The left (right) column is p+p (d+Au) data.
The curves are fits described in the text, including the area of the 
back-to-back peak ($S$).
\label{fig:corr}}
\end{figure}

Exploratory measurements of the azimuthal correlations between 
the forward $\pi^0$ and midrapidity $h^\pm$ are seen in 
Fig.~\ref{fig:corr} for p+p and d+Au collisions.
The leading charged particle (LCP) analysis picks the 
track at $|\eta_h|<0.75$ with the highest 
$p_T>0.5\,$GeV/c, and computes 
$\Delta\phi=\phi_{\pi^0}-\phi_{LCP}$ for each event.
The $\Delta\phi$ distributions are normalized by the number of
$\pi^0$ seen at $\langle \eta \rangle=4.00$.
Correlations near $\Delta\phi=0$ are not expected due to
the $\eta$ separation between the $\pi^0$ and the LCP.
The data are fit to a constant plus a Gaussian 
for the back-to-back peak centered at $\Delta\phi=\pi$.
The fit parameters are correlated, and their
errors are from the full error matrix.
The values do not depend on $N_\gamma$.
The area $S$ under the back-to-back peak is the 
probability that a LCP is correlated with a forward $\pi^0$.
The area $B$ under the constant represents 
the underlying event.
The total coincidence probability per trigger $\pi^0$
is $S+B\approx 0.62\ (0.90)$ 
for p+p (d+Au), and is constant with $E_\pi$.
The ratio $S/B$ for p+p does not depend on midrapidity
track multiplicity.
The peak width has contributions from 
transverse momentum in hadronization and from momentum 
imbalance between the scattered partons.
                                                                               
A PYTHIA simulation \cite{pythia} including detector 
resolution and efficiencies predicts most features of the p+p 
data \cite{dis2004}.
PYTHIA expects $S \approx 0.12$ and $B \approx 0.46$, 
with the back-to-back peak arising from $2\rightarrow 2$ 
scattering, resulting in forward and midrapidity partons that 
fragment into the $\pi^0$ and LCP, respectively.
The width of the peak is smaller in PYTHIA than in the data,
which may be in part because the predicted momentum imbalance 
between the partons is too small, as was seen for back-to-back 
jets at the Tevatron \cite{d0}.

The back-to-back peak is significantly smaller in d+Au collisions
than in p+p, qualitatively consistent with 
the monojet picture arising in the coherent
scattering\,\cite{coherent} and CGC\,\cite{monojet} models.
HIJING \cite{hijing} includes a model of shadowing for nuclear 
PDFs. 
It predicts that the back-to-back peak in d+Au collisions should
be similar to p+p, with $S\approx 0.08$.
The data are not consistent with the HIJING expectation at low 
$E_\pi$.
                                                                               
In conclusion, the inclusive yields of forward $\pi^0$ mesons from 
p+p collisions at $\sqrt{s}=200\,$GeV generally agree with NLO pQCD
calculations.
However, by $\langle\eta\rangle=4.00$, the spectrum is found to 
be harder than NLO pQCD, becoming suppressed with decreasing $p_T$.
In d+Au collisions, the yield per binary collision is suppressed 
with increasing $\eta$, decreasing to $\sim 30\%$ of 
the p+p yield at $\langle\eta\rangle=4.00$, well below shadowing 
and multiple scattering expectations, as well as exhibiting isospin 
effects at these kinematics.
The $p_T$ dependence of the d+Au yield is consistent with a model 
which treats the Au nucleus as a CGC.
Exploratory measurements of azimuthal correlations of the forward 
$\pi^0$ with charged hadrons at midrapidity show a recoil peak in 
p+p collisions that is suppressed in d+Au at low $E_\pi$, as would 
be expected for monojet production.
These effects are qualitatively consistent with a gluon
saturation picture of the Au nucleus,
but cannot definitively rule out other 
interpretations.
A systematic program of measurements, including direct 
photons and di-hadron correlations over a broad range of 
$\Delta\eta$, $p_T$, and $\sqrt{s}$, is 
needed to explore the nuclear modifications to particle production.
A quantitative theoretical understanding of the observables 
is needed to facilitate experimental tests of a possible color 
glass condensate.
        
We thank the RHIC Operations Group and RCF at BNL, and the
NERSC Center at LBNL for their support. This work was supported
in part by the HENP Divisions of the Office of Science of the 
U.S.\ DOE; the U.S.\ NSF; the BMBF of Germany; IN2P3, RA, RPL, 
and EMN of France; EPSRC of the United Kingdom; FAPESP of Brazil;
the Russian Ministry of Science and Technology; the Ministry of
Education and the NNSFC of China; IRP and GA of the Czech 
Republic, FOM of the Netherlands, DAE, DST, and CSIR of the 
Government of India; Swiss NSF; the Polish State Committee for 
Scientific Research; STAA of Slovakia, and the Korea Sci.\ \& 
Eng.\ Foundation.
Support of the RIKEN-BNL Research Center for one of the authors
(BF) to participate in this work is gratefully acknowledged.

\end{document}